\documentclass[twocolumn,amssymb, nobibnotes, showpacs, superscriptaddress, aps, prd]{revtex4}
\usepackage{amsmath}

\begin{document}
\title{Superluminal propagation of an optical pulse in a Doppler broadened 
three-state, single channel active Raman gain medium}
\author{K.J Jiang}
 \affiliation{Electron \& Optical Physics
Division, NIST, Gaithersburg, Maryland 20899}
\affiliation{Center for Cold Atom Physics,
Wuhan Institute of Physics and Mathematics, CAS, Wuhan, 200062, China}
\author{L. Deng}
 \affiliation{Electron \& Optical Physics
Division, NIST, Gaithersburg, Maryland 20899}
\author{M.G. Payne}
 \affiliation{Electron \& Optical Physics
Division, NIST, Gaithersburg, Maryland 20899}

\date{\today}

\begin{abstract}
Using a single channel active Raman gain medium we show a $(220\pm 20)$ns advance time for 
an optical pulse of $\tau_{FWHM}=15.4\,\mu$s propagating through a 10 cm medium,
a lead time that is comparable to what was reported previously.  In addition, 
we have verified 
experimentally all the features associated with this single channel Raman gain 
system. Our results show that the reported gain-assisted superluminal propagation 
should not be attributed to the interference between the two frequencies of the pump field.  
\end{abstract}

\maketitle
\noindent
In a recent work \cite{lit1} in room-temperature Cs atom and with two nearby
Raman resonances created by two pump frequencies, a $3.7\,\mu$s optical pulse
is observed to travel superluminally with about 62 ns lead time, corresponding 
to about 1.6 \% of the full width at half maximum (FWHM) of the probe pulse 
used in the experiment.  In their conclusion the authors have attributed the 
observed superluminal propagation to the interference effect between 
different frequency components of the pump field  \cite{lit1a} (in the anomalous 
dispersion region provided by the two resonances) and also to the gain that 
occurs at the frequency of the probe pulse. Payne et al. \cite{lit2} have 
shown analytically, however, that in this case the superluminal propagation is
not based on any interference effect.  Indeed, they have shown that
with a pump field of single frequency, the superluminal propagation 
should still be present.  In this work, we demonstrate experimentally 
that an active Raman gain medium enabled by a {\em single} frequency pump field 
can indeed bring about the superluminal propagation, as predicted in Ref.\cite{lit2}.  
Thus, while the attribution of the observed superluminal propagation is 
assisted by Raman gain is correct it is incorrect to attribute the origin of
the superluminal propagation to an interference between the two pump frequency 
components.  It is the gain nature of the anomalous dispersion that is at the 
root of the apparent superluminal propagation observed.
\vskip 10pt

\noindent
Before describing our experiments we first cite several pioneer works on 
superluminal propagations of optical pulses in anomalous dispersion
regions of optical media.  This include the early work of Garrett and McCumber
\cite{lit3} on the 
propagation of a Gaussian light pulse through an anomalous dispersive medium 
, the study by Casperson and Yariv \cite{lit4} on pulse proration in a high-gain
medium, the work by Chu and Wong \cite{lit5}on the linear pulse propagation 
in an absorbing medium.  More recently, Chiao \cite{lit6} investigated
the superluminal propagation of wavepackets in transparent media with
inverted atomic population, Bolda et al. \cite{lit7} studied optical pulse 
propagation with negative group velocity due to a nearby gain line, and Steinburg 
et al. \cite{lit8,lit9} investigated single photon tunneling time and superluminal 
propagation in a medium with a gain doublet. Stenner et al. \cite{lit10} have studied the
information speed in superluminal propagation.
\vskip 10pt

\noindent
We use $^{85}$Rb to demonstrate the superluminal propagation of an optical pulse
in a single frequency channel active Raman gain medium (Fig. 1).  An external cavity diode 
laser system produces nearly 400 mW power and is 
locked to a rubidium reference cell.  The output of the laser passes through 
a 3 GHz acousto-optical modulator (AOM) to generate two 
laser beams with frequency difference of the ground state hyper-fine levels $|1\rangle=|5S_{1/2}, F=2\rangle$ and $|3\rangle=|5S_{1/2}, F=3\rangle$ of 
$^{85}$Rb atoms.  The AOM is driven by a RF synthesizer which 
allows accurate adjustment so that the two-photon resonance can be precisely 
maintained.  A frequency shifter is then added to the path of the probe field,
allowing accurate and independent tuning of the small two-photon detuning.
Figure 1 shows the energy level diagram and relevant laser 
couplings for the present experiment.  The probe and the pump fields couple
the $|1\rangle=|5S_{1/2}, F=2\rangle\rightarrow|2\rangle=|5P_{1/2}\rangle$ and
$|3\rangle=|5S_{1/2}, F=3\rangle\rightarrow|2\rangle=|5P_{1/2}\rangle$ transitions,
respectively.  Both lasers are detuned, however, on the high 
energy side of the $5P_{1/2}$ hyper fine manifold by a large ($\simeq$3GHz)
one-photon detuning.  In this experiment the strong pump field serves two purposes: (1)
it is on the $|1\rangle=|5S_{1/2}, F=2\rangle\rightarrow |2\rangle=|5P_{1/2}\rangle$ resonance (see discussion below), and (2) it also pumps
the $|3\rangle=|5S_{1/2}, F=3\rangle\rightarrow |2\rangle=|5P_{1/2}\rangle$
transitions with a large one-photon detuning.  The former serves as the
optical pumping field that keeps nearly all population in the state $|3\rangle$,
whereas the latter provides the active Raman gain to a weak probe field tuned near
the $|1\rangle=|5S_{1/2}, F=2\rangle\rightarrow|2\rangle=|5P_{1/2}\rangle$
transition.  It is in this active Raman gain region where we study the superluminal
propagation of the weak probe pulse.  Contrary to the two-frequency pump field
scheme used in the ref. \cite{lit1}, there is no second pump frequency component
that can lead to the interference effect discussed in \cite{lit1}.
Finally, the cell for the medium 
is 100 mm in length and 25 mm in diameter with anti-reflection coating on both 
ends, and is placed in a temperature
controlled enclosure.  Experiments were carried out in the temperature region of 
$T=60\,^{\text{o}}$C to $T=90\,^{\text{o}}$C.
\vskip 10pt

\noindent
Theoretically, the system described above can be understood using a simple 
life time broadened three-state
model where nearly all population remains in the initial state $|3\rangle$.  
This is because under our experimental conditions the one-photon 
detuning is sufficiently large in comparison with the Doppler
broadened line widths and the ac Stark shift produced
in the $5P_{1/2}$ hyperfine states by the optical pumping field.
Consequently, the Doppler broadenings of the hyperfine states are un-important
and the contributions by the two hyperfine states can be easily included.  
Thus, we neglect the 
optical pumping field, the Doppler broadening, and the hyperfine splitting, and assume
that nearly all population is maintained in the 
state $|3\rangle$.  Since the probe field is very
weak \cite{lit11,lit12}, there is never appreciable population can be transfered 
to the state 
$|1\rangle$.  
\vskip 10pt

\noindent
The equations of motion for the relevant density matrix elements 
are given as (assuming $\rho_{33}\approx 1$)
\begin{subequations}
\begin{eqnarray}
\frac{\partial{\rho_{12}}}{\partial{t}}&\approx&-i\Omega_c^*e^{i\Delta_c t}\rho_{13}-\gamma_{12}\rho_{12},\\
\frac{\partial{\rho_{13}}}{\partial{t}}&\approx&i\Omega_{p}^*e^{i\Delta_p t}\rho_{23}-i\Omega_{c}e^{-i\Delta_c t}\rho_{12}-\gamma_{13}\rho_{13},\\
\frac{\partial{\rho_{23}}}{\partial{t}}&\approx&i\Omega_{p}e^{-i\Delta_p t}\rho_{13}+i\Omega_{c}e^{-i\Delta_c t}-\gamma_{23}\rho_{23},
\end{eqnarray}
\end{subequations}
where $\Omega_{p}=D_{21}E_{p0}/(2\hbar)$ and $\Omega_{c}=D_{23}E_{c0}/(2\hbar)$ 
are the half Rabi frequencies of the
probe ($E_{p0}$) and pump ($E_{c0}$) fields for the respective transitions, $D_{ij}$ and 
$\gamma_{ij}$ are the electric dipole moment and the decoherence rate of the 
respective transitions.  In addition,
$\Delta_p$ and $\Delta_c$ are one-photon detunings to the state $|2\rangle$ 
by the probe and pump fields.
\vskip 10pt

\noindent
Equation (1) must be solved self-consistently together with the wave equation
for the pulsed probe field in order to correctly predict the propagation 
characteristics of the probe field.
Payne et al. have shown that the positive frequency part of the probe field 
amplitude, in the case of un-focused plane wave and within the slowly varying 
amplitude and adiabatic approximations, must satisfy the wave equation which,
in the Fourier transform space, is given by
\begin{subequations}
\begin{eqnarray}
\frac{\partial{\Lambda_{p}}}{\partial{z}}-i\frac{\omega}{c}\Lambda_{p}=\frac{i\kappa_{12}|\Omega_c|^2W(\omega)}{(\Delta_c-i\gamma_{23})(\Delta_c+i\gamma_{21})}\Lambda_p,\\
W(\omega)=\frac{1}{\omega-(\Delta_c-\Delta_p-i\gamma_{31})-\frac{|\Omega_c|^2}{(\Delta_c+i\gamma_{21})}},
\end{eqnarray}
\end{subequations}
where $\Lambda_{p}$ is the Fourier transform of the probe field Rabi frequency
$\Omega_p$ and $\omega$ is the Fourier transform variable.
In addition, $\kappa_{12}=2\pi N_0\omega_{p}|D_{21}|^2/c$ where $N_0$ and $\omega_{p}$
are the atom number density and frequency of the probe field, respectively.
We note that the last term in the denominator in Eq. (2b) leads to a small
shift of the Raman gain peak
and a shift of the minimum group velocity.  
This term has the origin in the small ac Stark shift due to the pump field. 
\vskip 10pt

\noindent  
Payne et al. have carried out a non-steady state calculation and shown the group
velocity of a probe pulse as (for the two-photon detuning $|\delta_{2ph}|=|\Delta_p-\Delta_c|\gg\gamma_{31}$),
\begin{equation}
V_g\approx -\frac{\Delta_c^2\left(\delta_{2ph}-\frac{|\Omega_c|^2}{\Delta_c}\right)^2}{\kappa_{12}|\Omega_c|^2},
\end{equation}
where $|\Delta_p|,|\Delta_c|\gg\gamma_{21}$ (assuming $\gamma_{21}\approx\gamma_{23}$),
and $|\Delta_p|,|\Delta_c|\gg|\Omega_c|$ have been used.  We note that the negative 
sign indicates the superluminal propagation, thus proving the 
interference  between the two-frequency components of a pump field is not the 
cause of the superluminal propagation.  It is the active Raman gain that is 
responsible for apparent superluminal propagation.  
\vskip 10pt

\noindent
Our experiment is 
aimed at demonstrating three predictions given in Eq. (3):  (1) the existence of 
the non-distorted apparent superluminal propagation in a single pump
frequency channel, active Raman gain medium; (2) the characteristic inverse 
parabolic dependence of the group velocity on the pump field Rabi frequency; 
and (3) quadratic dependence of the group velocity on the two-photon detuning
(neglecting the small ac Stark shift).
\vskip 10pt

\noindent
In Fig. 2 we show a typical data of the advanced propagation of a Gaussian probe 
pulse with FWHM pulse width of $\tau_{FWHM}=15.4\,\mu$s.  
In order to show the 
advanced time clearly we have plotted only a small portion of the probe pulse profile.
The lower-left panel shows the advance of the front
edge (solid line) in comparison with a reference Gaussian pulse (dashed line) that
traverses through the air, whereas the lower-right panel depicts the advance of the 
rare edge in comparison with the same reference pulse. 
We have observed excellent S/N ratio \cite{lit13} and the fitting of the data 
to a Gaussian using a standard statistical routine has yielded
the advanced time of $\delta t=(220\pm 20)$ ns, or about 1.4 \% of the FWHM pulse
width of the Gaussian probe pulse.  This is comparable to the advance time of the
two-pump-frequency experiment where 1.6 \% advanced time has been reported \cite{lit1}.
\vskip 10pt

\noindent
In Fig. 3a we show the group velocity of a Gaussian probe pulse as a function 
of the single frequency pump field Rabi frequency.  The inverse quadratic 
behavior (notice the negative sign) of the group velocity as the function of the
pump field Rabi frequency is in accord to the prediction based on Eq. (3) when the
pump field Rabi frequency is relative small therefore the ac Stark shift can be
neglected.  As the pump field Rabi frequency is strong, the ac Stark shift term in
Eq. (3) must be considered together with the ground state population depletion.
In particular, a sizable ac Stark shift leads to the transition from the inverse 
quadratic behavior to the quadratic increase of magnitude of the group velocity as
can be seen from Eq. (3) at large $\Omega_c$ limit.  In this limit, however, one
should re-derive Eqs. (1-3) by including the equation of motion for the ground state 
population and the possible effects due to the ``accidental" optical pumping and a 
four-wave mixing (FWM) field generated by the allowed transition $|2\rangle\rightarrow|3\rangle$ due to the ``accidental" optical pumping field.  
This complicated situation shall be studied later.
\vskip 10pt

\noindent
In Fig. 3b we show the group velocity of a probe pulse as a function of 
the two-photon detuning $\delta_{2ph}=\Delta_p-\Delta_c$.  Equation (3)
predicts that the group velocity is a quadratic function of the two-photon
detuning.  In addition, when $|\delta_{2ph}|\gg Max[|\Omega_c|^2/|\Delta_c|,\gamma_{31}]$, the group velocity should be 
symmetric with respect to the two-photon detuning.  This is indeed what
has been observed experimentally. Here, we plot the red two-photon detuning
portion of the group velocity dispersion curve.   
It is worth noting that the vertical dashed line indicates
the approximate two-photon detuning where the propagation characteristics
of the probe pulse propagation changes from subluminal (above the horizontal
dashed line) to superluminal (below the horizontal dashed line).
This behavior is in complete agreement with the predictions based on Eq. (2b).
Indeed, from Eq. (2b) it is seen that the when $\delta_{2ph}\approx|\Omega_c|^2/\Delta_c$,
the $i\gamma_{31}$ term dominates the denominator \cite{lit14}, resulting in 
a subluminal propagation.  This point occurs at about, for red detuned two-photon
detuning, 
\begin{equation}
\delta_{2ph}\approx\frac{|\Omega_c|^2}{\Delta_c}\approx -2\pi\times 200\,\text{kHz}.\nonumber
\end{equation}
This is very close to what can be seen from Fig. 3b.  
\vskip 10pt

\noindent
It is also worth pointing out another observation that is in accord with the prediction of
Eq. (2a,2b) and ref. \cite{lit2}.  We have observed about 1\% {\em narrowing} 
of the probe pulse for $|\delta_{2ph}|\le 2\pi\times$ 400 kHz.  Superluminal propagation
without pulse narrowing, such as those depicted in Fig. 2 are observed 
for $|\delta_{2ph}|>2\pi\times$ 400 kHz, as expected from Eq. (3) and ref.\cite{lit2}.
With a shorter probe pulse such as $\tau_{FWHM}=5\,\mu$s, we have observed nearly 10 \%
pulse narrowing in addition to a $(295\pm 20)$ ns lead time.  This lead time is about 5 \% 
of the FWHM width of the probe pulse used.
\vskip 10pt

\noindent
We now discuss the possible effect of the ``accidental" optical pumping field
and the likely hood of the possible FWM generation due to the allowed
transition $|2\rangle\rightarrow|3\rangle$ because of the presence of the ``accidental" 
optical pumping field.  In our simple three-state treatment given
before, the optical pumping field is not included.  Consequently, the possible
generation of a FWM field is also neglected.  We neglect the ``accidental" 
optical pumping field and the FWM field in our treatment because (1).	theoretically 
with all four fields included, one could not find a clean analytical solution 
to the problem, therefore the extract of the essential physics is non-trivial.  
In addition, steady-state solution is highly questionable.  
This is because the equations of motion contain fast oscillating factors 
that cannot be removed by simple phase transformation; (2).	although the "accidental" 
optical pumping field is on resonance with the $|1\rangle\rightarrow|2\rangle$ 
transition, and therefore causes perturbation to the dispersion properties of 
state $|2\rangle$, this perturbation is not very significant to the gain process
and the propagation of the probe field simply because the detunings from state 
$|2\rangle$ are so large.  It should be pointed out that the contribution 
by FWM is small in our case even with a sizable gain. Theoretically, however,
the gain increases with the pump field intensity, and the ground state depletion 
will occur at sufficient gain.  In this strong pumping regime, the inclusion of 
the "accidental" optical pumping field and the FWM generation becomez necessary 
in order to accurately predict the propagation dynamics.
\vskip 10pt

\noindent
In conclusion we have verified experimentally all predictions based on Eq. (3)
and ref. \cite{lit2}.
We have demonstrated experimentally that the superluminal propagation of an
optical pulse in an active Raman gain medium is {\em not} the result of the interference
between two frequency components in the pump field.  Indeed, with only a single
frequency pump field where no second frequency component exists to allow the 
possibilty of interference between the pump frequencies but with nearly the same
Raman gain coefficient ($G_{Raman}\simeq 0.05$ cm$^{-1}$ in our exoeriment and 
$G\simeq 0.04$ cm$^{-1}$ reported in Ref. \cite{lit1}) we have shown  
comparable advance times as reported before by other group.  In addition, 
we have demonstrated four features
of the single channel active Raman gain medium for superluminal propagation:
(1) the inverse quadratic dependence of the group velocity as a function of
the pump field Rabi frequency; (2) the quadratic and symmetric
dependence of the group velocity
as a function of the two-photon detuning; (3) the group velocity dispersion
region where the propagation characteristics changes from superluminal to subluminal;
and (4) probe pulse narrowing when $|\delta_{2ph}|<\gamma_{31}$.
None of these features has been demonstrated before with a narrowband gain meidium.

\newpage
Figure captions
\vskip 10pt
{\bf{Figure 1}}  Top panel: Energy level diagram and relevant laser couplings. 
Level assignment:
$|1\rangle=|5S_{1/2}, F=2\rangle$, $|2\rangle=|5P_{1/2}$,
and $|3\rangle=|5S_{1/2}, F=3\rangle$.  Because
of the large one-photon detuning, both the hyper fine splitting,
Doppler broadening and small ac. Stark shift induced in the state $|2\rangle$
have been neglected. Lower panel: Schematics of the experiment set up.
\vskip 10pt
{\bf{Figure 2}} Typical data (solid line) showing the advanced propagation of a 
Gaussian probe pulse of full width at half maximum (FWHM) $\tau_{FWHM}=15.4\,\mu$s 
(top panel, the dashed line is for a reference pulse).
Here, we have plotted portions of the front (lower left) 
and rare (lower right)
edges of the probe pulse (solid line) in comparison with a reference Gaussian pulse 
that travels in the air (dashed line).  The vertical axis is the normalized optical 
intensity of the Gaussian probe pulse.  The probe pulse has a slightly
higher intensity because
of the Raman gain.  There is, however, no detectable pulse spreading or distortion 
when the probe field amplitude is normalized. Parameters: $|\Omega_c|=2\pi\times 25$ MHz,
$\delta_{2ph}\approx 400$ kHz, $T=60\,^{\text{o}}$C, and $\Delta_c=2\pi\times 3$ GHz.
\vskip 10pt
{\bf{Figure 3}} (a) Plot of the group velocity of a Gaussian probe pulse as a function 
of the pump field Rabi frequency (a).  As expected, the group velocity exhibits
the inverse quadratic dependence of the Rabi frequency of the single channel 
pump field, as predicted in Eq. (3). For this plot $\delta_{2ph}\approx 2\pi\times 400$ kHz,
$\Delta_c=2\pi\times 2.2$ GHz, and temperature $T=60\,^{\text{o}}$C.
(b) Plot of the group velocity of a Gaussian probe pulse as a function 
of the two-photon detuning $\delta_{2ph}$.  In the superluminal region where
$|\delta_{2ph}|\gg\gamma_{31}$, the group velocity exhibits the quadratic 
dependence on the two-photon detuning, as 
predicted from Eq. (3).  When $|\delta_{2ph}-|\Omega_c|^2/\Delta_c|<<\gamma_{31}$
the propagation characteristics changes from superluminal to subluminal.
For this plot $|\Omega_c|=2\pi\times 25$ MHz,
$\Delta_c=2\pi\times 3$ GHz, and temperature $T=60\,^{\text{o}}$C. The Raman gain 
coefficient of $G_{Raman}\simeq 0.05$ cm$^{-1}$ used in our exoeriment is similar to
the gain coefficeint of $G\simeq 0.04$ cm$^{-1}$ reported in Ref. \cite{lit1}.


\begin{thebibliography}{13}
\bibitem{lit13}{In our experiment, we choose the Raman gain coefficient to be about
$G_{Raman}\approx 0.05\,\text{cm}^{-1}$.  This is comparable to the gain coefficient
of $G\approx 0.04\,\text{cm}^{-1}$ reported in Ref.[1]}
\bibitem{lit14}{That is, when the denominator in Eq. (2b) changes form a mostly real to
mostly imaginary, the propagation characteristics changes from superluminal to subluminal.}
\end{thebibliography}
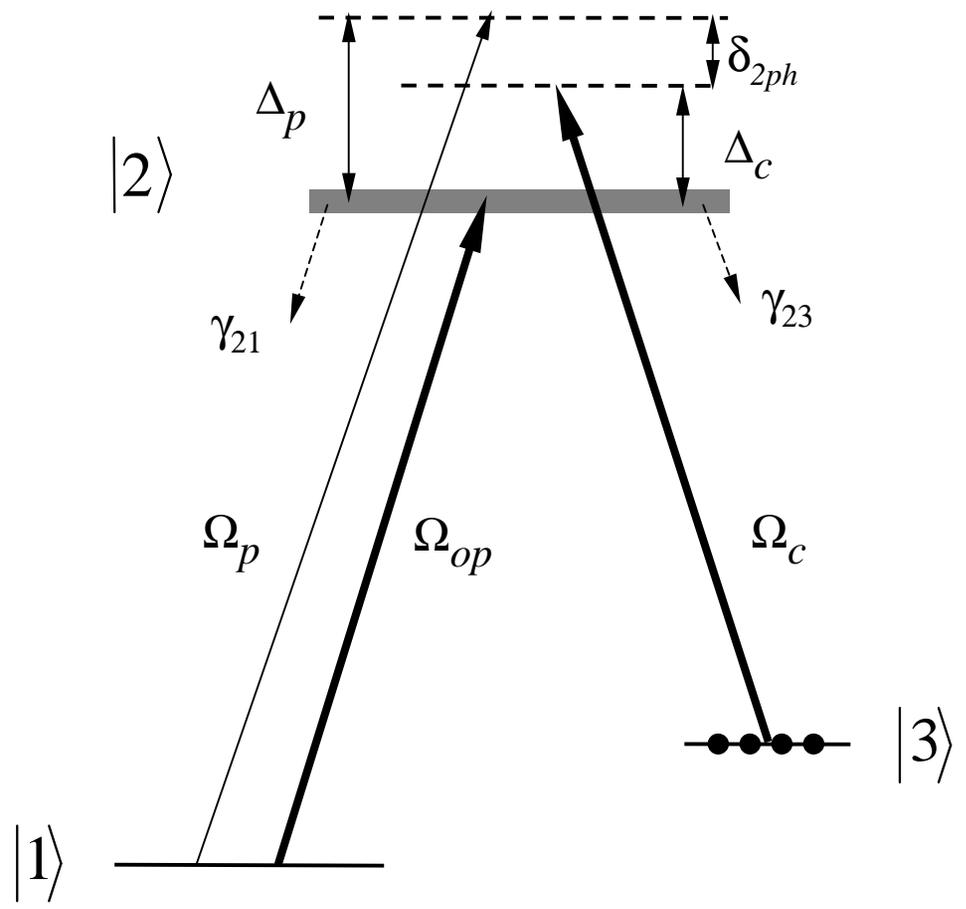
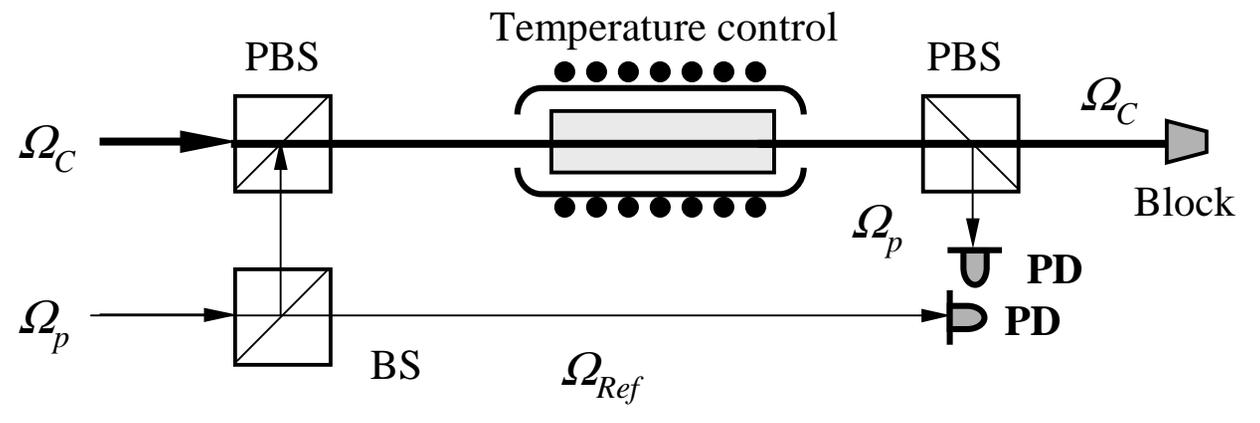
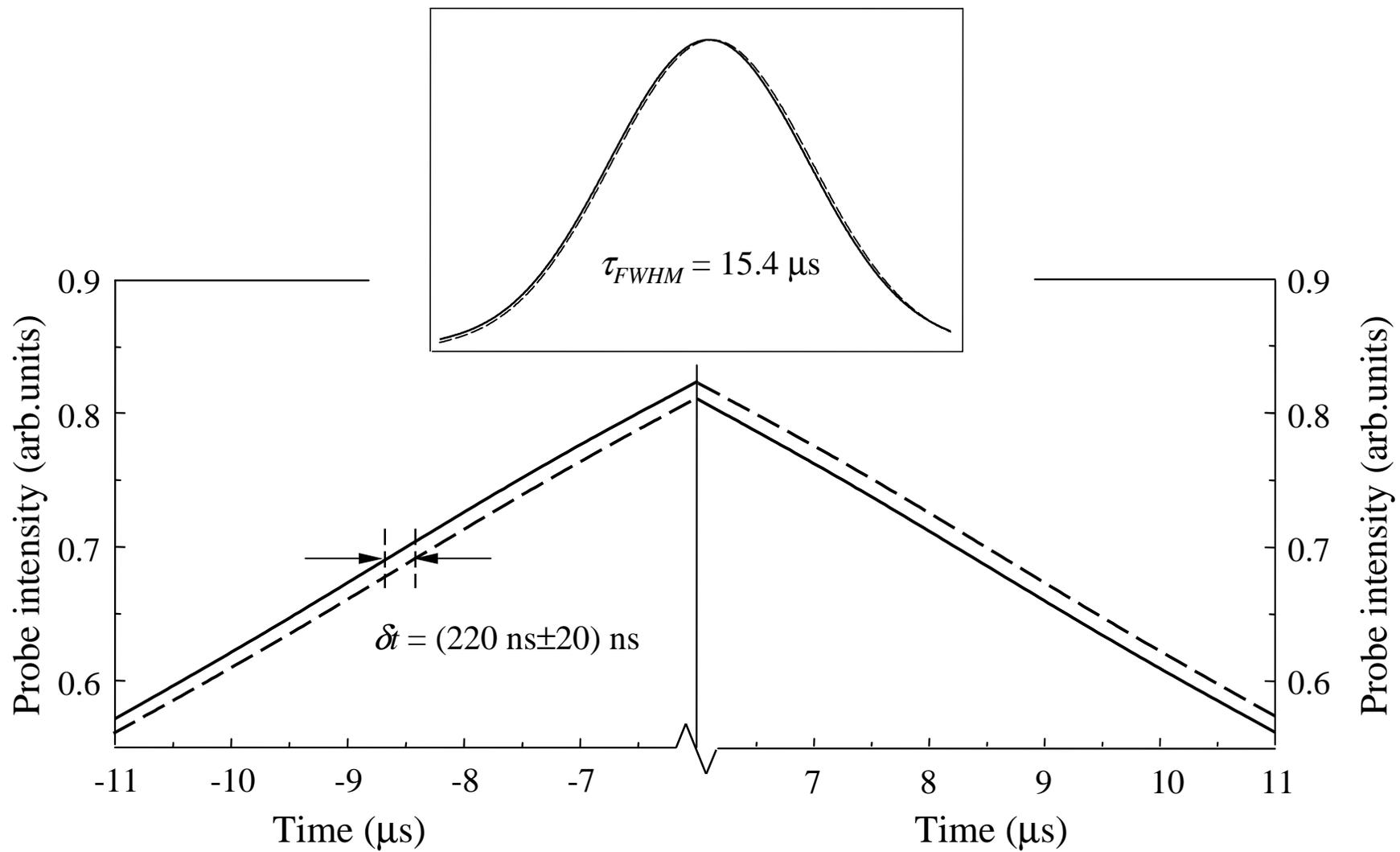
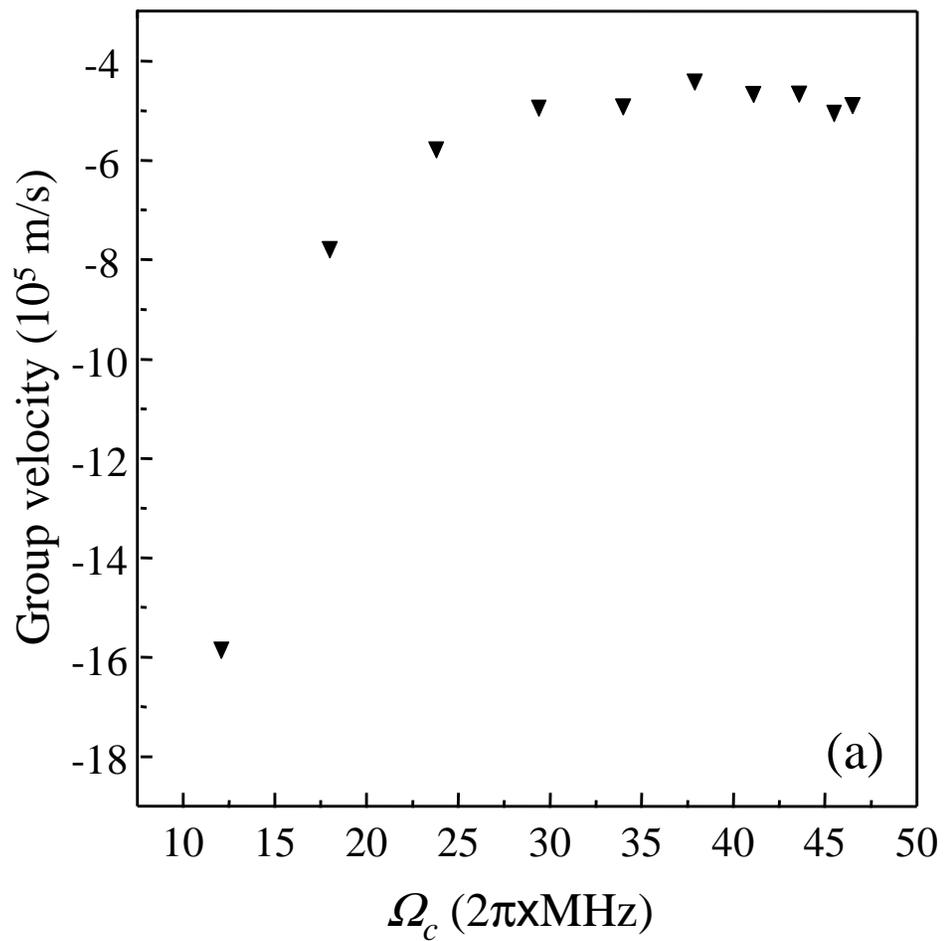
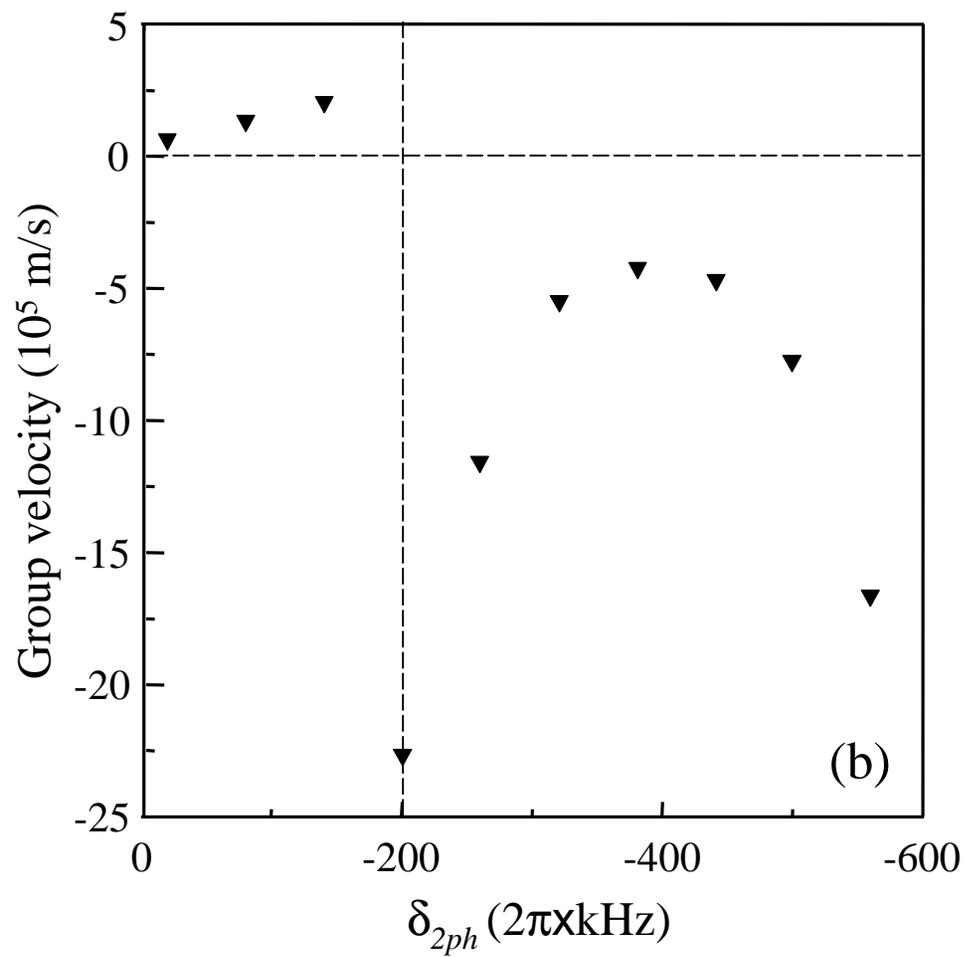
\end{document}